**Chemical Shift Encoding based Double Bonds Quantification in Triglycerides using Deep Image Prior**

Chaoxing Huang,[1,2] Ziqiang Yu,[1,2] Zijian Gao,[1,2], Qiuyi Shen[1,2], Queenie Chan[3], Vincent Wai-Sun Wong[4], Winnie Chiu-Wing Chu[1,2], Weitian Chen [1,2]

1. Department of Imaging and Interventional Radiology, The Chinese University of Hong Kong, Hong Kong SAR, China.
2. CUHK Lab of AI in Radiology (CLAIR), The Chinese University of Hong Kong, Hong Kong SAR, China.
3. Philips Healthcare, Hong Kong SAR, China.
4. Department of Medicine & Therapeutics, State Key Laboratory of Digestive Disease, The Chinese University of Hong Kong, Hong Kong SAR, China.

Address correspondence to:

Weitian Chen, PhD

Department of Imaging and Interventional Radiology, Prince of Wales Hospital, The Chinese University of Hong Kong, 30-32 Ngan Shing Street, Shatin, New Territories, Hong Kong SAR, China

CUHK Lab of AI in Radiology (CLAIR), Prince of Wales Hospital, The Chinese University of Hong Kong, 30-32 Ngan Shing Street, Shatin, New Territories, Hong Kong SAR, China

(852)-3505-1036

Email: wtchen@cuhk.edu.hk

Numbers of Figures and Tables: 7

Running Title: DIP for Mapping Double Bonds

**Abstract**

Fatty acid can potentially serve as biomarker for evaluating metabolic disorder and inflammation condition, and quantifying the double bonds is the key for revealing fatty acid information. This study presents an assessment of a deep learning approach utilizing Deep Image Prior (DIP) for the quantification of double bonds and methylene-interrupted double bonds of triglyceride derived from chemical-shift encoded multi-echo gradient echo images, all achieved without the necessity for network training. The methodology implemented a cost function grounded in signal constraints to continually refine the neural network's parameters on a single slice of images through iterative processes. Validation procedures encompassed both phantom experiments and in-vivo scans. The outcomes evidenced a concordance between the quantified values and the established reference standards, notably exemplified by a Pearson correlation coefficient of 0.96 ($p = 0.0005$) derived from the phantom experiments. The results in water-oil phantom also demonstrate the quantification reliability of the DIP method under the condition of having a relatively low-fat signal. Furthermore, the in-vivo assessments showcased the method's competency by showcasing consistent quantification results that closely mirrored previously published findings concerning subcutaneous fat. In summary, the study underscores the potential of Deep Image Prior in enabling the quantification of double bonds and methylene-interrupted double bonds from chemical-shift encoded multi-echo magnetic resonance imaging (MRI) data, suggesting potential avenues for future research and clinical applications in the field.

Keywords: Deep Learning, Number of Double Bonds, Triglyceride, magnetic resonance imaging (MRI)

## 1. INTRODUCTION

Chemical-shift encoding–based water–fat separation methods have been developed to quantify fat content (1,2). Recently, there has been growing interest in quantifying the fatty acid composition of fat due to its potential in evaluating metabolic disorders and inflammatory conditions (3). Previous studies indicate that the key to quantifying the fatty acid composition lies in determining the number of double bonds in triglycerides (4-7). Optimization algorithms have been proposed to quantify this variable using chemical-shift encoded multi-echo methods (4,5,8).

In recent years, there is strong interest in applying deep neural networks in quantitative magnetic resonance imaging (qMRI) for tasks such as image reconstruction and parameter mapping as they are more robust compared to conventional fitting methods(9). Quantitative MRI collects data in higher dimensions than the conventional anatomical imaging. Deep learning shows promising performance in handling high-dimensional data (9). However, most deep learning-based methods require a substantial amount of data to train the neural network. For many medical imaging tasks, it is challenging or even impractical to collect sufficient amount of training data. Deep image prior (DIP) method (10) has been proposed as an unsupervised learning method to address ill-posed inverse problems using deep network without requiring any training data. The original DIP approach was found promising in image denoising and restoration tasks (10-12). The recent work demonstrated the potential of DIP in medical imaging tasks where training data is difficult to acquire, including PET image reconstruction and restoration (13-15) and MR image reconstruction (16-19).

One challenge for quantifying the number of double bonds in triglycerides is that it is highly susceptible to signal perturbations, making it an ill-posed problem. Deep learning approaches were previously reported promising in quantitative MRI in the presence of signal perturbations (20,21). However, training a deep neural network requires a sufficient amount of data and the data available for quantifying fatty acid is scarce in the community. The properties of DIP make it a proper choice to address these issues in quantifying the numbers of double bonds in triglycerides via neural networks. In this work, we investigated DIP for mapping the number of double bonds in triglycerides from multi-echo MRI. We demonstrated the feasibility using both phantom and in vivo experiments.

## 2. METHODS

### 2.1 Signal model

The multi-peak multi-echo signal with water and fat contents at echo time $t$ can be expressed as (4) :

$$S(W,F,ndb,nmidb,\phi;t) = (W + Ff\sum_{m=1}^{M}\alpha_m(ndb,nmidb)e^{i\omega_m t})e^{\phi t} \qquad [1]$$

Where W, F are water and fat signal, respectively; $\phi$ is a complex map with the real and imaginary component representing the sum of field map and R2* map.; $f = 1/\sum_{m=1}^{M}\alpha_m$ is the normalization factor; $\alpha_m$ is the amplitude of the $m$th fat peak and is the function of the number of double bonds (ndb ) and the number of methylene-interrupted double bonds (nmidb); and $\omega_m$ is the known chemical shift of the $m$th fat peak. We adopt the eight-peak fat model in (4), which is also referred as the "free model". In free model, ndb and nmidb are estimated as independent parameters. More details of the free model can be found in Table 1 in (4).

### 2.2 Network fitting

We first recap the basic concept of deep image prior and present its extension to the mapping of numbers of double bonds in triglyceride. The DIP method is unsupervised and interprets the output of a deep network as a parametrization of an image (10). In the original DIP method, the deep network $f_\theta(\bullet)$ with network weight $\theta$ takes a randomized noise map $\mathbf{z}$ as input and map it to a denoised image with the same size. Given a noisy image $I$ , the denoised image $I^*$ can be obtained by minimizing the following equation:

$$\theta^* = \arg\min_{\theta} || I - f_{\theta}(\mathbf{z}) ||, I^* = f_{\theta^*}(\mathbf{z}) \quad [2]$$

The network is enforced to reconstruct the noisy image $I$ through iterations and early stopping is applied so that the output tensor is a denoised image rather than the noisy one. The DIP is based on an assumption that the structure of an autoencoder- like architecture already captures the main image statistics, which is independent of learning. Those statistics priors are associated with low-level image features, and it is the must-need information to recover an image with high fidelity from a degraded one. Intuitively, the clean image with high fidelity can emerge during the process of reconstructing the noisy image by using those image statistics.

As for quantifying the parametric maps in our task, the network is defined as a parametric map generator and the optimization can be defined as follows:

$$\tilde{\theta} = \arg\min_{\theta} \frac{1}{PE} \sum_{p=1}^{P} \sum_{e=1}^{E} [L(S(f_{\theta}(\mathbf{z}), t_e)_p, y_{ep})]$$

$$A = f_{\tilde{\theta}}(\mathbf{z}) \quad [3]$$

where $A$ is the estimated parametric map; $e$ is the index of the echo and $y_e$ stands for the acquired signal under the corresponding time of echo; $p$ is the index of the pixels in an image; and $L$ is the cost function and L1 norm is chosen as the cost function.

The network takes the randomized map as input and the network parameters are updated iteratively following the optimization in equation (3). The result arrives once the loss function is converged, and early stopping is applied. There is no training for the neural network as the parameters of the network are updated solely for the MR parameters of a single slice, and the optimization itself can be regarded as the inference. The network has five outputs, including W, Ff, $\phi$, ndb and nmidb . The complex images W, Ff, $\phi$ have two output channels, one for the real part and the other for the imaginary part. The ndb and nmidb outputs have one channel. We group the fat signal term F and the normalization factor term f into one term as we empirically find it more stable to fit the network. The fat signal can be recovered once ndb and nmidb are obtained as they can be used to compute the normalization factor. The fat fraction (FF) is defined as $\mathrm{Re}al(F/(W+F))$ . Sigmoid function is placed at the channel of ndb to limit its range from 0 to 6 as this is the typical range of the numbers of double bonds (5).We choose not to limit the range of nmidb as we empirically found that it did not create much difference to the convergence. We adopt a UNet like architecture for the deep network (22). The optimization process is shown in Figure 1.

**2.3 Data Acquisition**

The in-vivo study was conducted in accordance with the Declaration of Helsinki (as revised in 2013). The study was approved by ethics review board at the Chinese University of Hong Kong (2016.150) and informed consent was taken from all individual participants.

We employed the multi-echo gradient echo sequence protocol described in (23) to acquire the images. Scans were performed using a Philips Elition 3T MRI scanner (Philips Healthcare, Best, the Netherlands). A 32-channel head coil and a 32-channel cardiac coil were used as the receiver for phantom and abdominal imaging, respectively. A total of 14 echoes were acquired with the following parameter settings: $TE_1/\Delta TE$ = 1.20/0.7 ms, TR = 10 ms, Flip angle = 20 degrees, matrix size = 160 × 120, FOV = 400 × 300 mm, and slice thickness = 6 mm. Seven slices were acquired. Frequency encoding direction is from right to left. All data were collected in axial plane. For in vivo scan, the data of each slice was acquired in a single breath hold of 17 seconds.

### 2.4 Implementation details

The experiments were conducted in a computing environment running Windows 11 with Python 3.7. All experiments were implemented using PyTorch 1.9 (24). The computational tasks were performed on a system equipped with an RTX 4090 GPU (24GB) and an i9-13900 CPU. For comparative analysis, we also implemented the algorithm described in (4) to quantify the number of double bonds by fitting the data to the eight-peak free model. In the subsequent discussion, we refer to this method as least square fitting (LSF), as it is based on the LSF method of water-fat separation (2). All images were cropped to ensure the samples occupied the majority part of the image. The Adadelta optimizer (25) was employed, and the learning rate was set to 5e-2. The fitting of DIP was conducted with 180,000 iterations and it took around an hour to complete. The fitting of the LSF took 15 to 30 minutes in our experiments.

### 2.5 Phantom experiment

To validate the measured $ndb$ values of different vegetable oils by comparing them to the literature values previously published (5,26), we prepared 50mL tubes of various kinds of pure vegetable oil (Olive, Peanut, Safflower, Walnut, Grapeseed, Canola, and Corn oil). These tubes were immersed in water bath in our phantom. In addition to the pure oil phantoms, we also used corn oil and created a representative water-fat phantom with a fat fraction of approximately 60% to evaluate the algorithm's ability to quantify the number of double bonds in the presence of mixed water and fat. The creation of the water-fat phantom followed the protocol outlined in (27). The agar solution was prepared by heating and mixing distilled water, agar powder, sodium dodecyl sulfate, and sodium benzoate (all sourced from Sigma Aldrich, St. Louis, MO, USA). The agar and oil solutions were blended, stirred, heated, and subsequently cooled to form the water-fat phantom.

The analysis of the results was performed within regions of interest (ROIs). Circular ROIs were placed at the center of each tube. The mean and the standard deviation of all pixel values within the ROIs of all slices were calculated.

### 2.6 In-vivo experiment

We performed an in-vivo abdomen scan on a healthy volunteer. The ROIs were chosen on the area of subcutaneous fat. The mean value within the ROIs of all the slices was compared with previously published literature values (5,26).

## 3. Results

Table 1 shows the fitted result compared with the literature values of both the DIP and the LSF method.

Figure 2 shows the maps of the number of double bonds and the number of methylene-interrupted double bonds from the pure oil phantoms. Note the measured values are close to the expected values, and all the tubes show an FF of nearly 100%.

Figure 3 illustrates the regression plot depicting the correlation between the measured values and the literature values of ndb and nmidb for both the DIP method and the LSF method. The plot reveals a strong correlation, and the regression lines closely align with the reference plot of y=x. For the LSF method, the Pearson correlation coefficient between the measured values and the literature values is 0.98 ($p$ = 8.89e-5) for ndb and 0.98($p$ = 9.34e-5) for nmidb. The regression analysis yields a slope of 1.04 and an intercept of -0.29 for ndb and a slope of 0.98 and an intercept of -0.04 for nmidb . Similarly, the DIP method exhibits a reasonable correlation performance, with a Pearson correlation coefficient of 0.96 ($p$ = .0005) for ndb and 0.96 ($p$ = .0006) for nmidb . The regression analysis of DIP yields a slope of 0.92 and an intercept of 0.35 for ndb and a slope of 0.90 and an intercept of 0.20 for nmidb .

Figure 4 shows the results from the phantom with mixed water and fat made from corn oil. The DIP method yields consistent measurement in two phantoms (pure oil ndb = 4.50±0.21, mixed phantom ndb = 4.43±0.24, pure oil nmidb = 1.88±0.12, mixed phantom nmidb = 1.83±0.15). The LSF method shows inconsistent measurement in two phantoms (pure oil ndb = 4.32±0.22,mixed phantom ndb = 5.27±0.26, pure oil nmidb = 1.72±0.13,mixed phantom nmidb = 2.58±0.16), respectively. The robustness of DIP method can also be appreciated from its closer fat fraction (FF) result (62.76%) to the reference value (60%). In contrast, the LSF method yields FF with a larger deviation (66.70%). For DIP, there is no significant differences between the result of pure oil and mixed phantom ($p$=0.20 for ndb and $p$=0.15 for nmidb).

Figure 5 shows the in-vivo results obtained from a typical slice using both methods. Note the published reference value of ndb and nmidb of fat in the subcutaneous region are 2.88 and 0.70, respectively (5). The measured values of the healthy volunteer are shown in Table 2, indicating a reasonable alignment of the measured values and the literature values. It is also observed that there is left-right variation of the quantification map along the frequency encoding direction, likely due to echo shift along frequency encoding.

## 4. DISCUSSION

We demonstrated the feasibility of using the DIP-based method to measure the numbers of double bonds in triglycerides from chemical shift-encoded multi-echo gradient echo images. The results obtained from pure oil phantoms and subcutaneous fat show agreement between the measured values and the literature values. We observed that the DIP method demonstrated a superior performance compared to LSF method in the mixed water-fat phantom. We attribute the better performance to the denoising ability of the DIP method. It is likely that the LSF method requires exceedingly high fat fraction to provide sufficient SNR for reliable quantification of the number of double bonds. The convolutional neural network architecture itself is assumed to have the ability to capture implicit prior of the image, filter out noise and generate the main content of the image with high fidelity. In the original DIP method for denoising application, the network is enforced to reconstruct a noisy image while the image with reduced noise emerges after early-stop of iterations. In our work, the DIP network is enforced to reconstruct the gradient echo images from the output parametric maps. It is likely inherent denoising effect in the signal domain contribute to the fitting of the parametric maps, and thus improves robustness to estimate $ndb$ and $nmidb$ map compared to the LSF method when the fat signal is reduced.

It is important to note that the DIP-based method for fitting the numbers of double bonds is unsupervised and does not need pretraining the neural network. Self-supervised learning methods have been proposed for parametric MRI, which do not require ground truth for training but still require a substantial number of unlabeled images as the training data (28,29). In contrast, DIP only trains the network on a single dataset and the training itself is the inference. It may have potential significance in other learning-based qMRI mapping tasks, in which it is challenging to acquire training data.

While our initial validation of using DIP to fit the numbers of double bonds is promising, this work has limitations. The network needs to be retrained each time for a new mapping task, which was also pointed out in the original DIP work (10). This increases the computation cost of inference. New methods for accelerating the optimization process have been proposed recently (30) and it is worth looking into the application in qMRI. Quantifying fatty acid content is closely associated with water fat separation, and we may study the performance of the algorithms on more challenging scenarios that may produce water-fat swap and consider applying advanced correction method (31-34). In addition, the performance of the fitting algorithm can be studied on different water-fat models (7). Further work includes looking into the exact reason of left-right variation of in-vivo quantification maps and the corresponding measures to alleviate those impact. The impact of complex field map needs to be explored in the future as $R_2^*$ and field inhomogeneity may provide more clinical information. This study uses measurements from previous publication as reference value for validation. It is worthy of conducting gas chromatography and use it as a reference value in future studies. It is also important to conduct more in vivo studies on patients with various fat fractions to understand the feasibility of the DIP approach in clinical imaging.

## 5. CONCLUSIONS

We demonstrated the feasibility of using DIP to fit the number of double bonds in triglycerides from chemical shift-encoded multi-echo gradient echo MRI. Further studies are needed to validate this approach on more types of oils and a larger cohort of subjects.

**Acknowledgments**

Funding: This study was supported by grants from the Research Grants Council of the Hong Kong SAR (Project GRF 14213322), and the Innovation and Technology Commission of the Hong Kong SAR (Project No. MRP/046/20x).

**Footnote**

Conflicts of Interest: All authors have completed the ICMJE uniform disclosure form. Queenie Chan serves as an unpaid editorial board member of *Quantitative Imaging in Medicine and Surgery* and is the employee of Philips Electronics Hong Kong Limited. Vincent Wai Sun Wong served as a consultant or advisory board member for AbbVie, AstraZeneca, Boehringer Ingelheim, Echosens, Gilead Sciences, Intercept, Inventiva, Merck, Novo Nordisk, Pfizer, Sagimet Biosciences, TARGET PharmaSolutions, and Visirna; and a speaker for Abbott, AbbVie, Echosens, Gilead Sciences, Novo Nordisk, and Unilab. He has received a grant from Gilead Sciences, and is a cofounder and a shareholder of Illuminatio Medical Technology. Weitian Chen is a co-founder and a shareholder of Illuminatio Medical Technology Limited and PrivacyPro Medical Technology Limited. The other authors have no conflicts of interest to declare.

Ethical Statement: The authors are accountable for all aspects of the work in ensuring that questions related to the accuracy or integrity of any part of the work are appropriately investigated and resolved. The study was conducted in accordance with the Declaration of Helsinki (as revised in 2013). The study was approved by ethics review board of the Chinese University of Hong Kong (No. 2016.150) and informed consent was taken from all individual participants.

## Figures

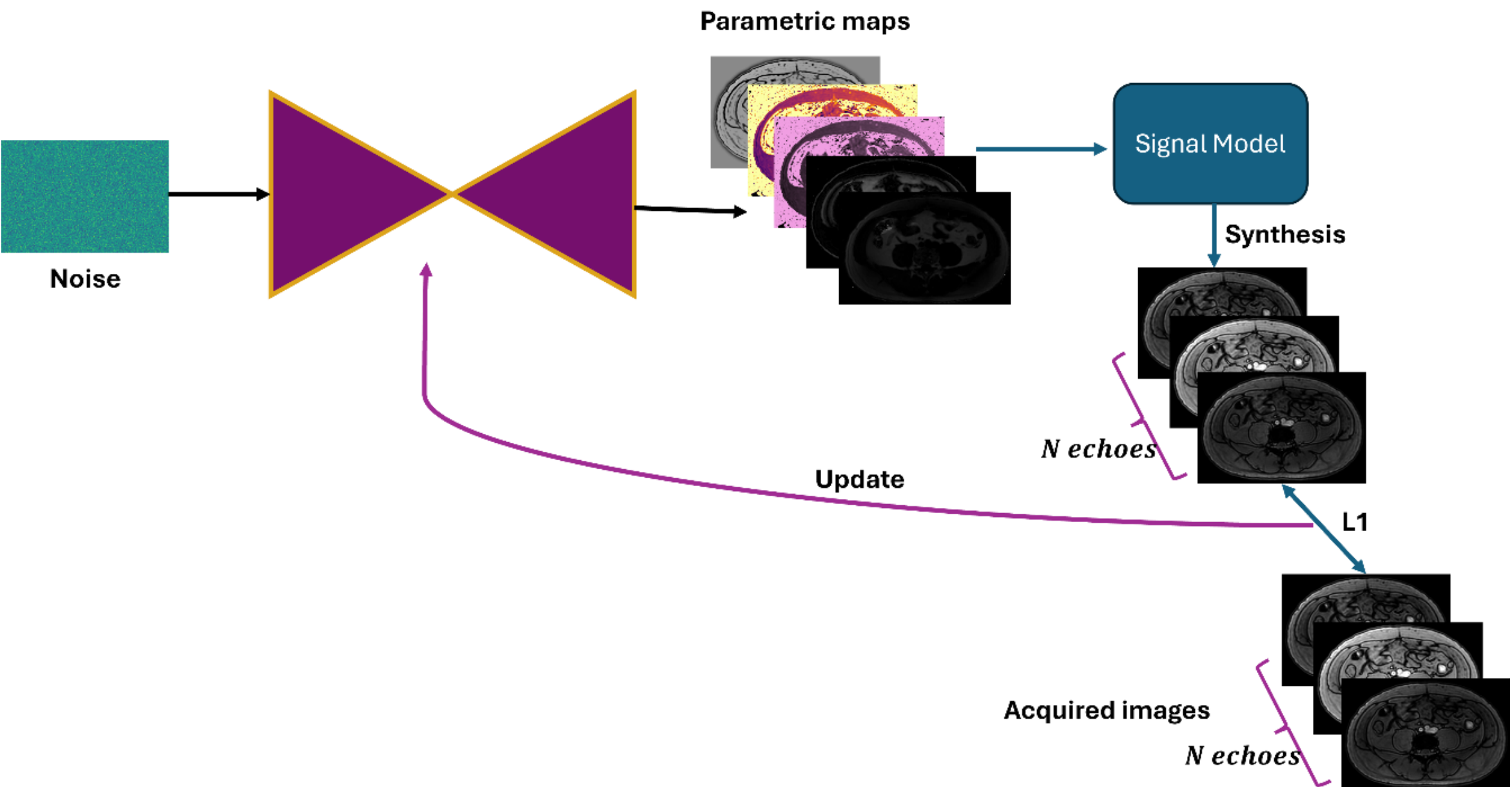

Figure 1: The optimization scheme of the network for fitting the parametric maps

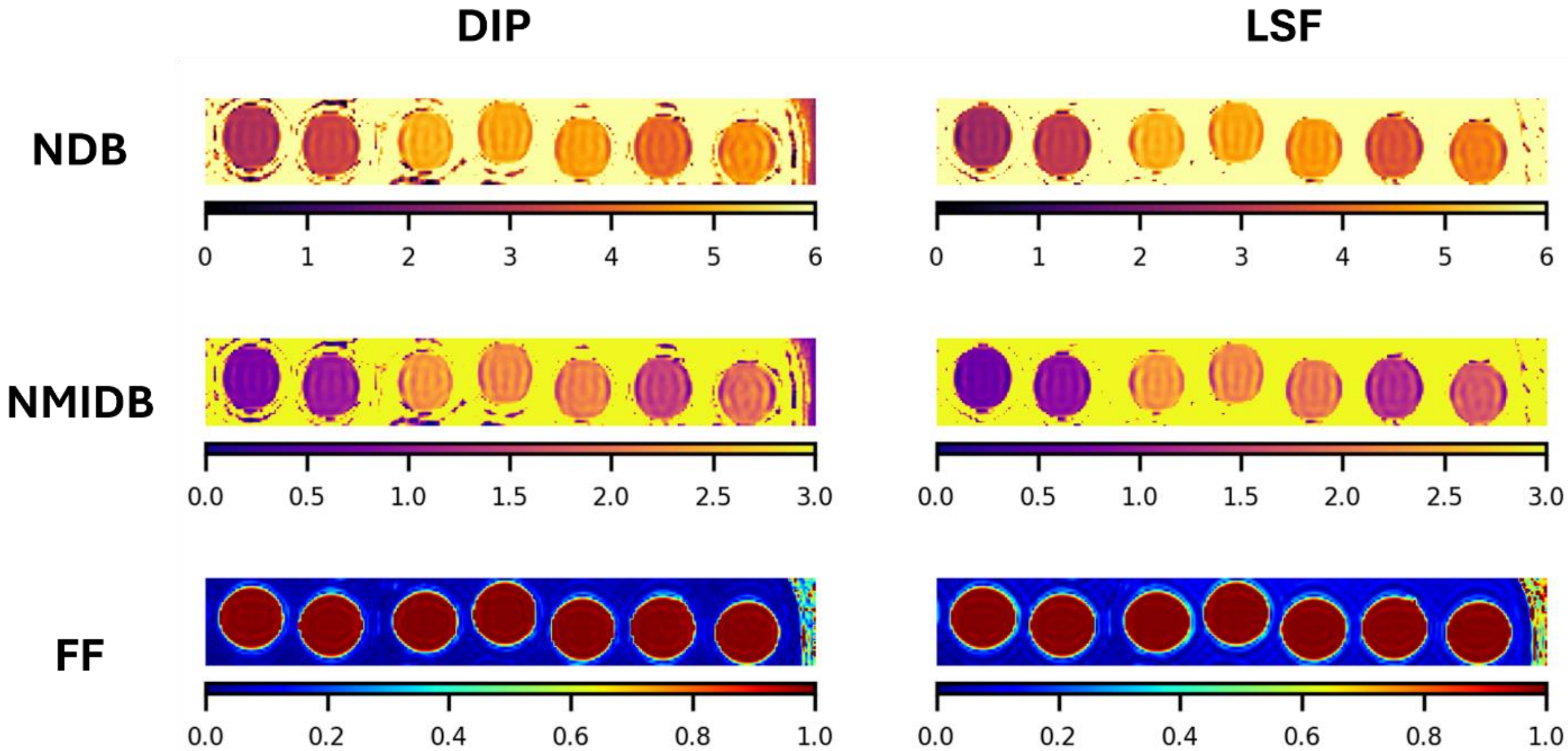

Figure 2: The measured $ndb$ and $nmidb$ using both the DIP and LSF method. From the left to the right in each plot are maps from Olive, Peanut, Safflower, Walnut, Grapeseed, Canola, and Corn oil, respectively. Abbreviations: DIP stands for Deep Image Prior, LSF stands for Least Square Fitting, NDB stands for number of double bonds, NMIDB stands for number of methylene-interrupted double bonds, FF stands for fat fraction.

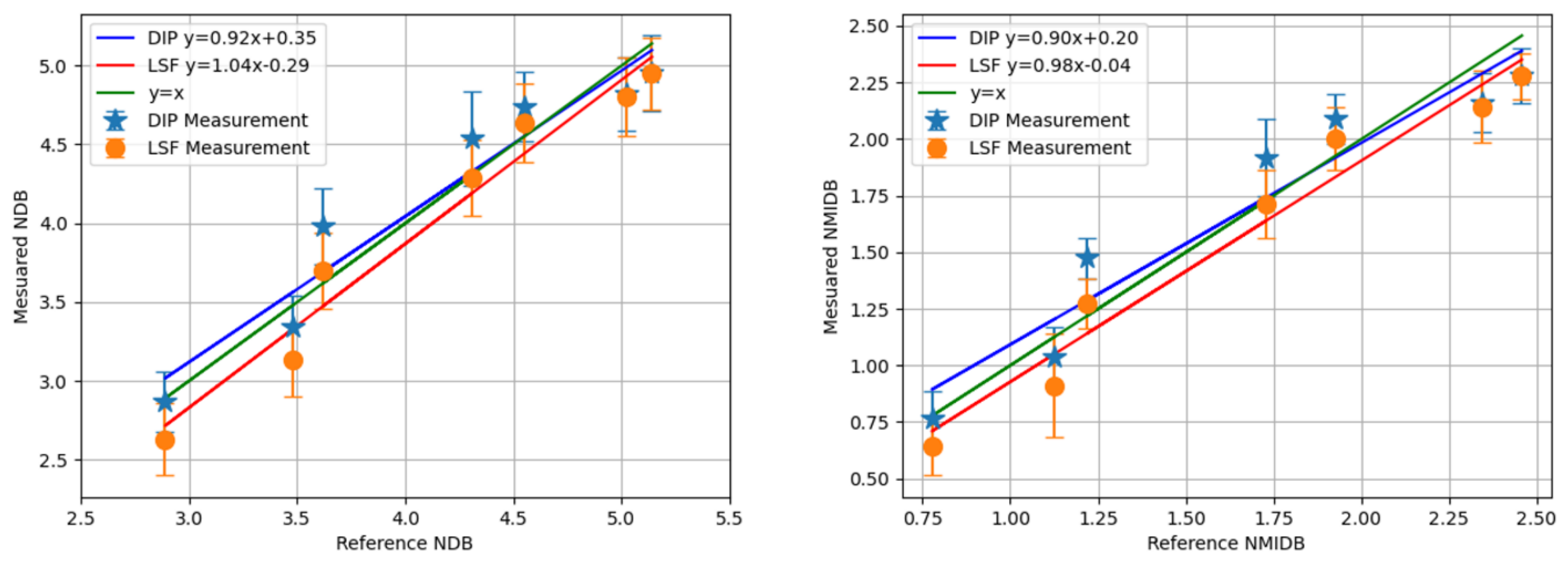

Figure 3: Regression plot of the measured $ndb$ and $nmidb$ values and the literature values. Abbreviations: DIP stands for Deep Image Prior, LSF stands for Least Square Fitting, NDB

stands for number of double bonds, NMIDB stands for number of methylene-interrupted double bonds

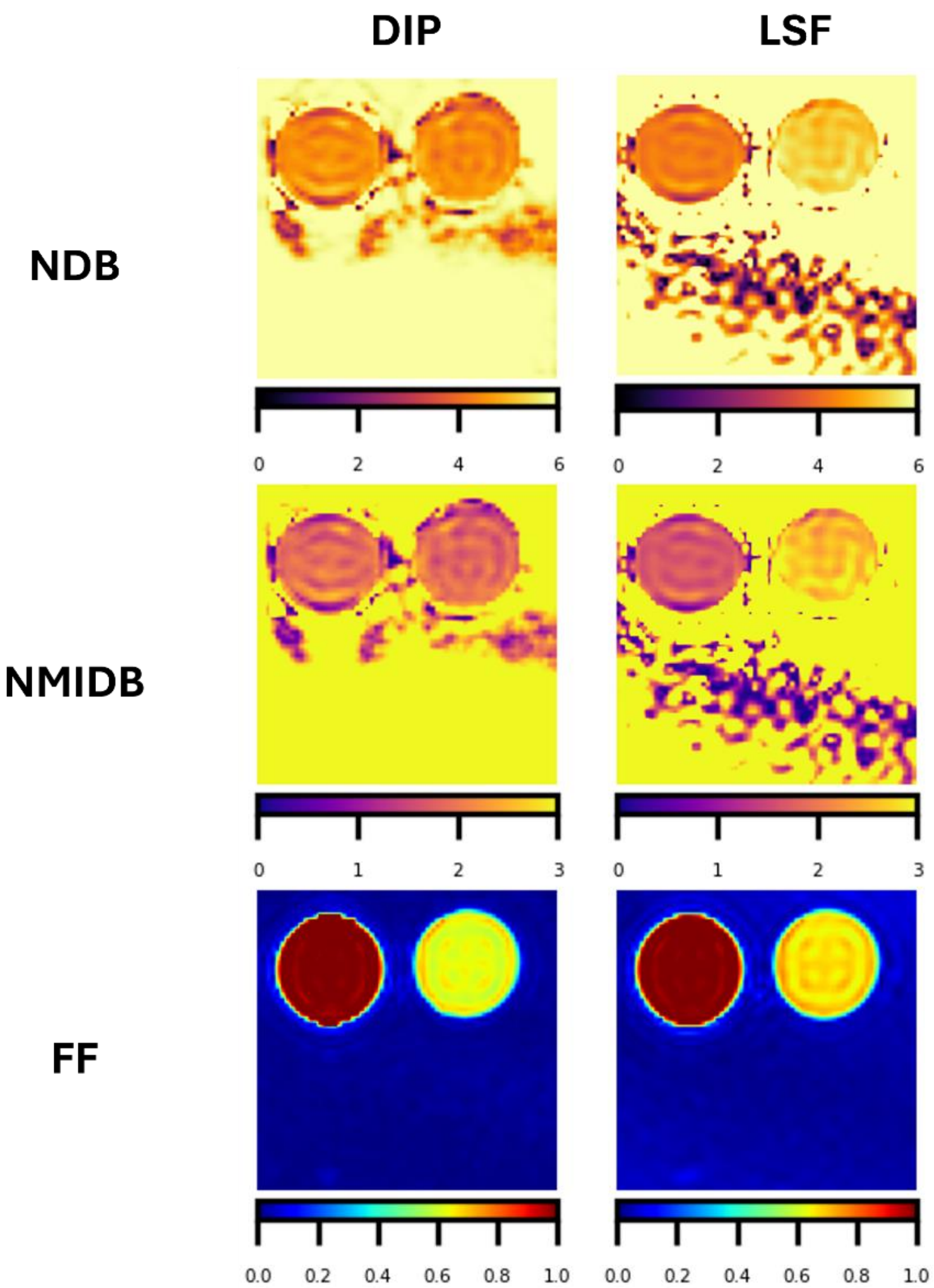

Figure 4: The measured number of double bonds from the pure corn oil phantom and the phantom with mixed water and corn oil. From the left to the right are the phantom with pure oil and the phantom with mixed oil and water (FF = 60%). Abbreviations: DIP stands for Deep Image Prior, LSF stands for Least Square Fitting, NDB stands for number of double bonds, NMIDB stands for number of methylene-interrupted double bonds, FF stands for fat fraction.

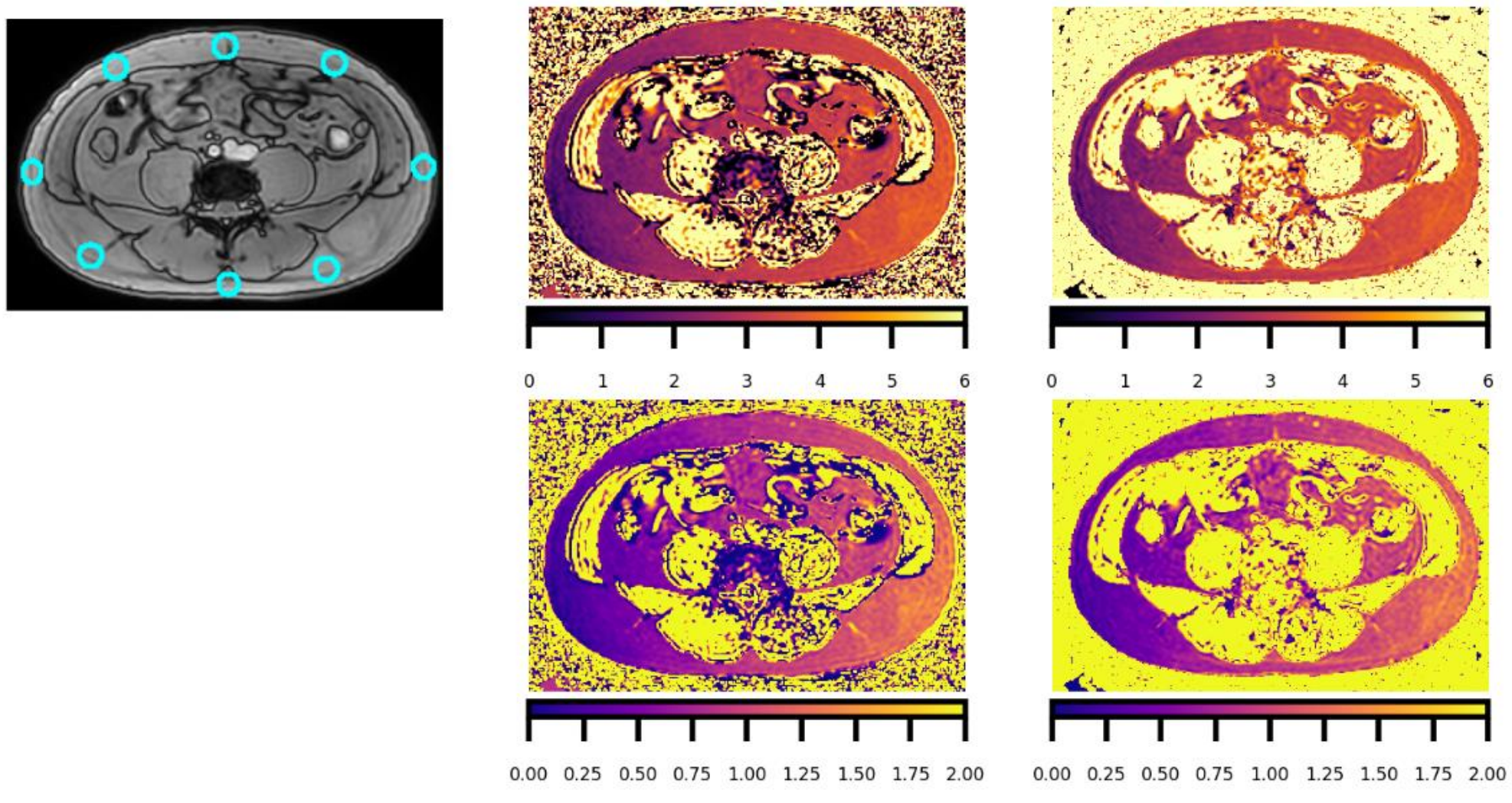

Figure 5: In-vivo mapping of $ndb$ and $nmidb$ using DIP and LSF. ROIS are indicated in light blue. Abbreviations: DIP stands for Deep Image Prior, LSF stands for Least Square Fitting, NDB stands for number of double bonds, NMIDB stands for number of methylene-interrupted double bonds, FF stands for fat fraction.

**Tables**

| Oil category | DIP | LSF | Literature values |
|---|---|---|---|
| **NDB** | | | |
| Safflower | 4.95±0.24 | 4.95±0.23 | 5.14 |
| Walnut | 4.82±0.23 | 4.80±0.25 | 5.02 |
| Grapeseed | 4.74±0.22 | 4.64±0.25 | 4.55 |
| Canola | 3.98±0.24 | 3.70±0.24 | 3.62 |
| Corn | 4.54±0.30 | 4.29±0.24 | 4.31 |
| Olive | 2.87±0.19 | 2.63±0.23 | 2.89 |
| Peanut | 3.34±0.20 | 3.13±0.23 | 3.48 |
| **NMIDB** | | | |
| Safflower | 2.28±0.12 | 2.27±0.10 | 2.34 |
| Walnut | 2.16±0.13 | 2.14±0.16 | 2.32 |
| Grapeseed | 2.08±0.11 | 2.00±0.14 | 2.3 |
| Canola | 1.47±0.09 | 1.27±0.11 | 1.14 |
| Corn | 1.92±0.17 | 1.71±0.15 | 1.75 |
| Olive | 0.77±0.12 | 0.64±0.13 | 0.35 |
| Peanut | 1.04±0.13 | 0.91±0.23 | 1.01 |

Table 1: The measured values of the NDB and NMIDB of different types of oils using DIP and LSF method from the pure oil phantoms. The reference value of the grapeseed oil is from reference (27) while the rest are from reference (5). DIP, Deep Image Prior; LSF, Least Square Fitting; NDB, number of double bonds; NMIDB, number of methylene-interrupted double bonds. The values are presented as mean±SD

|  | DIP | LSF |
| --- | --- | --- |
| ndb | 2.83±0.74 | 3.10±1.96 |
| nmidb | 0.74±0.35 | 0.89±0.15 |

Table 2: The measured NDB and NMIDB of the healthy volunteer for subcutaneous fat. DIP, Deep Image Prior; LSF, Least Square Fitting; NDB, number of double bonds; NMIDB, number of methylene-interrupted double bonds.